\documentclass[aps,twocolumn,prd,superscriptaddress,preprintnumbers,showpacs]{revtex4-1}
\usepackage{graphicx}
\usepackage{bm}
\usepackage{amsmath}
\usepackage{amssymb}
\usepackage{amsthm}

\newcommand{\be}{\begin{equation}}
\newcommand{\ee}{\end{equation}}
\newcommand{\bea}{\begin{eqnarray}}
\newcommand{\eea}{\end{eqnarray}}
\newcommand{\bdm}{\begin{displaymath}}
\newcommand{\edm}{\end{displaymath}}
\newcommand{\beas}{\begin{eqnarray*}}
\newcommand{\eeas}{\end{eqnarray*}}

\begin{document}
\title{Fine-structure constant constraints on Bekenstein-type models}

\author{P. M. M. Leal}
\email[]{up201002687@fc.up.pt}
\affiliation{Centro de Astrof\'{\i}sica, Universidade do Porto, Rua das Estrelas, 4150-762 Porto, Portugal}
\affiliation{Faculdade de Ci\^encias, Universidade do Porto, Rua do Campo Alegre, 4150-007 Porto, Portugal}
\author{C. J. A. P. Martins}
\email[]{Carlos.Martins@astro.up.pt}
\affiliation{Centro de Astrof\'{\i}sica, Universidade do Porto, Rua das Estrelas, 4150-762 Porto, Portugal}
\author{L. B. Ventura}
\email[]{up090322024@alunos.fc.up.pt}
\affiliation{Centro de Astrof\'{\i}sica, Universidade do Porto, Rua das Estrelas, 4150-762 Porto, Portugal}
\affiliation{Faculdade de Ci\^encias, Universidade do Porto, Rua do Campo Alegre, 4150-007 Porto, Portugal}
\date{4 June 2014}

\begin{abstract}
Astrophysical tests of the stability of dimensionless fundamental couplings, such as the fine-structure constant $\alpha$, are an area of much increased recent activity, following some indications of possible spacetime variations at the few parts per million level. Here we obtain updated constraints on the Bekenstein-Sandvik-Barrow-Magueijo model, which is arguably the simplest model allowing for $\alpha$ variations. Recent accurate spectroscopic measurements allow us to improve previous constraints by about an order of magnitude. We briefly comment on the dependence of the results on the data sample, as well as on the improvements expected from future facilities.
\end{abstract}

\maketitle

\section{\label{intro}Introduction} 

Experimental results from the LHC strongly suggest that fundamental scalar fields are among Nature's building blocks \cite{atlas,cms}. The recent indications from the BICEP2 experiment of primordial gravitational waves \cite{bicep2}, if confirmed, further suggest that these additional dynamical degree of freedom also played a crucial role on cosmological scales, at least in the first fraction of a second of the Universe. A pressing follow-up question is then whether astrophysical imprints of these additional degrees of freedom can also be detected in the more recent universe, since characterizing their behavior throughout the cosmological evolution would be an optimal way to distinguish among the extensive range of existing models.

Arguably the most obvious role for a dynamical scalar field in the recent universe would be as an alternative to a cosmological constant, in explaining the recent phase of accelerated expansion \cite{SN1,SN2}. But irrespective of whether or not this is the case, whenever dynamical scalar fields are present, one naturally expects them to couple to the rest of the model, unless a yet-unknown symmetry suppresses these couplings. In particular, a coupling to the electromagnetic sector will lead to spacetime variation of the fine-structure constant---see \cite{uzanLR} for a recent review.

Astrophysical tests of the stability of these couplings are therefore a powerful test of the consistency of the standard $\Lambda$CDM paradigm, as well as a direct probe of new physics beyond it. In recent years this has proven to be a very active area of observational research. In addition to the theoretical motivation to carry out these tests, there has been some recent evidence for such a variation \cite{webb}, and a significant effort is being put into independently confirming this result, including through a dedicated VLT/UVES Large Program \cite{LP1,LP2,bonifacio}.

In this brief report we use the most recent spectroscopic measurements of the fine-structure constant $\alpha$ to obtain constraints on the (arguably) simplest model where $\alpha$ variations can occur, the so-called Bekenstein-Sandvik-Barrow-Magueijo model \cite{bsbm,bsbm2} (which we will henceforth denote BSBM). This is a simple toy model where by construction the dynamical degree of freedom responsible for the $\alpha$ variation has a negligible effect on the cosmological dynamics. However, note that models also exist where this is not the case: an example is the string-inspired runaway dilaton scenario \cite{Dilaton1,Dilaton2,Dilaton3} (and other more phenomenological models are reviewed in \cite{uzanLR}).

Although one may argue that this BSBM is too simplistic, for our present purposes we take it simply as a phenomenological model to be constrained by data---with the advantage that since (to a good approximation) the $\alpha$ field does not affect the background dynamics we can assume that the standard cosmological parameters still apply and the model has a minimal number of additional free parameters. Thus we may simplistically envisage the BSBM model as $\Lambda$CDM plus a varying $\alpha$. Our goal here is therefore to explore how recent measurements improve constraints on this model, as compared to the original work of \cite{bsbm}. We will also discuss how the various current (sub)samples, which have somewhat different sensitivities and redshift distributions, constrain the model.

We note that extensions of the BSBM model with additional (functional) degrees of freedom appeared in recent years \cite{extend1,extend2}. Given the quantity and quality of the available data these do not have a strong observational motivation and we will not attempt to constrain them here. Neverthelesse, one may foresee that it will be possible to constrain such extendend theories though future, higher precision datasets.

\section{\label{model}The BSBM model}

The BSBM model was introduced in \cite{bsbm}, drawing on earlier work from Bekenstein \cite{bekenstein}. Here we provide a brief sumary of its relevant features, referring the reader to the original reference for a more detailed discussion as well as some additional motivation. Conceptually this is a dilaton-type model, but where the field is assumed to couple ony to the electromagnetic sector of the Lagrangian. The model's dynamical equations are obtained by standard variational principles.

Assuming a flat, homogeneous and isotropic cosmology, one obtains the following Friedmann equation
\be
\left(\frac{\dot a}{a}\right)^2=\frac{8\pi G}{3}\left[\rho_m(1+\omega\zeta e^{-2\psi})+\rho_r e^{-2\psi}+\rho_\Lambda+\frac{1}{2}\omega{\dot\psi}^2  \right]
\ee
and the scalar field equation is
\be
{\ddot\psi}+3H{\dot\psi}=-2\zeta\rho_m e^{-2\psi} \,.
\ee
Here $\omega$ is a parameter that can be defined as
$\omega\sim\hbar c/\ell^2$,
where $\ell$ effectively describes the scale below which one has significant deviations from standard electromagnetism. For simplicity (and consistently with \cite{bsbm}) we take $\omega\sim1$, leaving the coupling $\zeta$ as the only free parameter in the model. Typical values for this free parameter are discussed in some detail in \cite{bsbm,bsbm2}, but this discussion is not strictly relevant in the present context. For our purposes, irrespective of theoretical priors, the coupling prameter $\zeta$ is taken as a free phenomenological parameter, to be constrained by observations.

Note that in addition to radiation and matter the model needs a cosmological constant to match cosmological observations; the dynamical scalar field $\psi$ is subdominant in the dynamics of the universe (so we can assume the standard values of the cosmological parameters), and its only role is to drive a variation of the fine-structure constant. Specifically
$\alpha/\alpha_0= e^{2(\psi-\psi_0)}$ though the observational parameter of choice is the relative variation of $\alpha$, namely
\be
\frac{\Delta\alpha}{\alpha}(z)\equiv\frac{\alpha(z)-\alpha_0}{\alpha_0}\,;
\ee
thus a negative value corresponds to a smaller value of $\alpha$ in the past.

It's also interesting to note that in this model matter obeys the standard conservation equation
\be
{\dot\rho}_m+3H\rho_m=0
\ee
while that of radation is changed
\be
{\dot\rho}_r+4H\rho_r=2{\dot\psi}\rho_r\,.
\ee
As first pointed out in \cite{tasos}, this has the interesting consequence that the redshift dependence of the cosmic microwave background temperature will also change relative to the standard one, in a way that is not independent from the evolution of $\alpha$. Specifically one can write
\be
\frac{T(z)}{T_0}=(1+z)\left(\frac{\alpha(z)}{\alpha_0}\right)^{1/4}\sim(1+z)\left(1+\frac{\Delta\alpha}{\alpha}\right)\,;
\ee
although the current sensitivity of CMB temperature measurements at non-zero redshift is only at the percent level, some consistency tests of this type of relations may become possible with future facilities.

In the original \cite{bsbm} the authors suggested, based on a binned compilation of the $\alpha$ data then available, that a coupling $\zeta\sim {\rm few}\times 10^{-4}$ would provide a good fit to the data, and although a detailed statistical analysis was not reported an inspection of their fig. 2 would suggest that the sensitivity of constraints on $\zeta$ should be at the $10^{-4}$ level. In what follows we will discuss how this sensitivity has improved with present data.

\section{\label{results}Analysis}

High-resolution spectroscopic observations of absorption clouds along the line of sight of quasars currently provide the most precise astrophysical measurements of the fine-structure constant $\alpha$. (See \cite{bonifacio} for a brief overview of ongoing research.) In particular, there have been recent indications of spacetime variations of $\alpha$ at the level of a few parts per million, in the approximate redshift range $1<z<4$. The most recent such results are those of Webb \textit{et al.}  \cite{webb,webb2}, and we will use them here; note that they are made of two different subsample, one of which consists of measurements made with the HIRES spectrograph at the Keck telescope, while the other has measurements by the UVES spectrograph at ESO's VLT.

A possible cause for caution regarding the above measurements is that they come from archival data. Several efforts have been made to confirm this result through dedicated measurements. A summary of these new measurements is in Table \ref{table1}; the latest of these efforts is the ongoing Large Program at the VLT UVES \cite{bonifacio}. We will use both of these datasets in our analysis. Note that the more recent dataset has fewer data points (and a smaller redshift span) but smaller uncertainties, the reverse being true for the Webb \textit{et al.} data. (Some older measurements with larger uncertainties exist---cf. \cite{uzanLR}---but they would carry negligible weight in our statistical analysis, so they will not be included.)

\begin{table}
\begin{tabular}{|c|c|c|c|c|}
\hline
 Object & z & ${ \Delta\alpha}/{\alpha}$ & Spectrograph & Ref. \\ 
\hline
HE0515$-$4414 & 1.15 & $-0.1\pm1.8$ & UVES & \protect\cite{alphaMolaro} \\
\hline
HE0515$-$4414 & 1.15 & $0.5\pm2.4$ & HARPS/UVES & \protect\cite{alphaChand} \\
\hline
HE0001$-$2340 & 1.58 & $-1.5\pm2.6$ &  UVES & \protect\cite{alphaAgafonova}\\
\hline
HE2217$-$2818 & 1.69 & $1.3\pm2.6$ & UVES &  \protect\cite{LP1}\\
\hline
Q1101$-$264 & 1.84 & $5.7\pm2.7$ &  UVES & \protect\cite{alphaMolaro}\\
\hline
\end{tabular}
\caption{\label{table1}Available specific measurements of $\alpha$. Listed are, respectively, the object along each line of sight, the redshift of the absorber, the measurement in parts per million, the spectrograph, and the original reference.}
\end{table}

\begin{figure}
\includegraphics[scale=0.5]{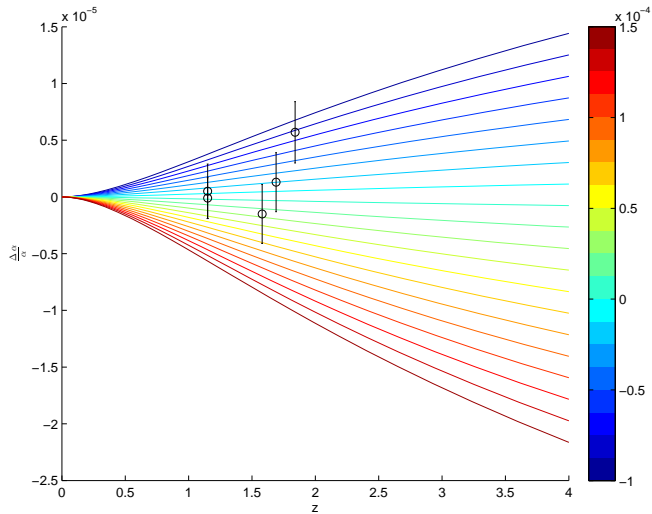}
\caption{The evolution of the fine-structure constant $\alpha$ (plotted as the relative variation with respect to the present laboratory value, $\Delta\alpha/\alpha$) as a function of the coupling parameter $\zeta$. The measurements listed in \protect\ref{table1} are also plotted.}
\label{fig1}
\end{figure}

Figure \ref{fig1} depicts redshift dependence of the fine-structure constant $\alpha$ in these models, as function of the coupling $\zeta$. The spectroscopic measurements in Table \ref{table1} are also plotted, to provide a simple illustration of the sensitivity of the observations to the coupling parameter. Notice that the $\alpha$ variation is always monotonic in this class of models. Also, the variation is quite small at low redshifts, and it also become significant deep on the matter era. Indeed, analytic studies of simple solutions of the model \cite{bsbm2} show that the matter-era asymptotic solution is
\be
\frac{\Delta\alpha}{\alpha}(z)\propto\ln{(1+z)}\,,
\ee
while once the universe starts to accelerate the evolution of $\alpha$ rapidly freezes. This feature is in fact essential for the model to fulfill local laboratory constraints, coming from atomic clock measurements \cite{Rosenband}.

\begin{figure}
\includegraphics[scale=0.5]{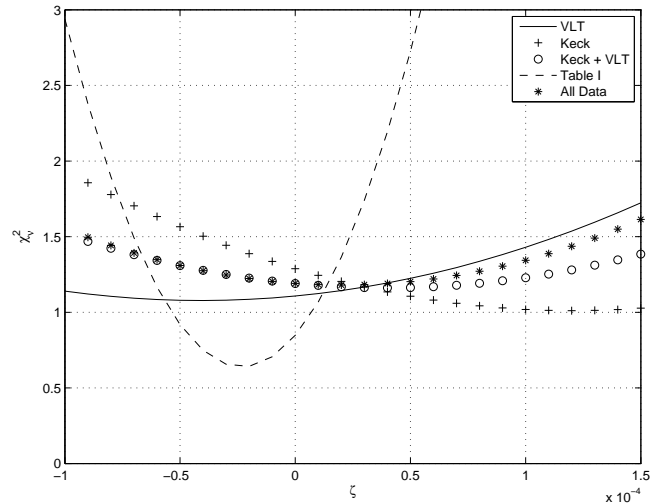}
\caption{Reduced chi-square for the sampled values of the coupling $\zeta$, with the dataset choices described in the text.}
\label{fig2}
\end{figure}

We then compare the model with the data and derive constraints on the coupling $\zeta$, though the standard chi-square statistic. Table \ref{table2} lists the one-sigma confidence intervals for $\zeta$, obtained for various choices of dataset, together with the corresponding reduced chi-square at the best-fit value. The values of $\chi^2$ in the relevant range of $\zeta$ are shown in Fig. \ref{fig2}.

\begin{table}
\begin{tabular}{|c|c|c|}
\hline
Data Sample & $\zeta$ & $\chi^2_{\nu,min}$ \\ 
\hline
Keck & $(12.1\pm1.9)\times10^{-5}$ & 1.01 \\
VLT & $(-4.1\pm1.9)\times10^{-5}$ & 1.08 \\
Keck+VLT & $(4.0\pm1.5)\times10^{-5}$ & 1.16 \\
\hline
Table \protect\ref{table1} & $(-2.3\pm2.9)\times10^{-5}$ & 0.64 \\
All & $(2.1\pm1.1)\times10^{-5}$ & 1.18 \\
\hline
\end{tabular}
\caption{\label{table2}One-sigma confidence interval for the coupling constant $\zeta$ for each analyzed dataset, together with the corresponding reduced chi-square at the best fit value.}
\end{table}

As is well known, both the Keck and VLT samples provide several-sigma indications of a varying $\alpha$, with opposite signs. This has been interpreted as evidence for a spatial dipole $\alpha$ \cite{webb,webb2}. We note that the BSBM is unable to account for parts per million level spatial variations (for a class of models that may conceivably do so see \cite{marvin}), so in the present context we take the data at face value and simply constrain variations with redshift. The Keck and VLT samples lead to preferred values of the coupling $\zeta$ with opposite signs, and more than two standard deviations away from zero. This is still the case for the combined (Keck $+$ VLT) sample; the former is statistically dominant, thus preferring a positive coupling, again at more than two standard deviations.

For the recent data in Table \ref{table1} there is no evidence for a non-zero coupling. Here the small reduced chi-square is noteworthy; one may speculate that this could be due to some of the error bars in the measurements being underestimated. In the case of the large program measurement \cite{LP1} the quoted error bar includes both a statystical and a systematic uncertainty (which have been added in quadrature), but whwther or not this has been done for other measureemnts is not always clear from the original references. 

If one combines the data of Webb \textit{et al.} with that of Table \ref{table1}, the evidence for a non-zero coupling decresases, and becomes less than two sigma. It's also noteworthy that the one sigma uncertaintly of the full dataset is significantly smaller than that of the Webb data, even though only five new measurements were added---the reason is, of course, that these are among the most precise measurements to date. This highlights the importance of target selection for the next generation of measurements.

\section{\label{concl}Conclusions}

We have used the results of recent astrophysical tests of the stability of the fine-structure constant $\alpha$ to we update constraints on the Bekenstein-Sandvik-Barrow-Magueijo model. At the phenomenological level this can be seen as a $\Lambda$CDM-like model with an additional dynamical degree of freedom which leads to a varying $\alpha$ without a significant impact on the Universe's dynamics, and in that sense it is the simplest model allowing for $\alpha$ variations.

In this brief report we have improved previous constraints by about an order of magnitude, but found no strong evidence for a non-zero coupling. However, one should kept in mind that the BSBM is in any case unable to account for a spatial variation of $\alpha$ at the parts per million level, as may be inferred from \cite{webb,webb2}.

Looking ahead, the ongoing UVES Large Program for Testing Fundamental Physics \cite{LP1,LP2,bonifacio} is trying to confirm the earlier evidence for variations, but is also playing a crucial role as preparatory work for the next generation of high-resolution ultra-stable spectrographs, such as ESPRESSO \cite{espresso} and ELT-HIRES \cite{hires} which have tests of the stability of fundamental couplings among their key science drivers. While the current sensitivity of individual measurements with UVES or similar spectrographs is at the 1-2 parts per million level, an improvement of about a factor of 3 is expected with ESPRESSO, and of at least an order of magnitude with ELT-HIRES.

Our results confirm the expectation that a small number of high precision measurements can have a significant impact in terms of constraining possible theoretical paradigms, and thereby highlight the importance of a theoretical input in the process of target selection \cite{obstrat} (in addition to the obvious observational practicalities). Untimately, in the E-ELT era, astrophysical tests of the stability of fundamental couplings will become a crucial part of a new generation of precision consistency tests of fundamental cosmology \cite{eelt}.

\begin{acknowledgments}
This work was done in the context of project PTDC/FIS/111725/2009 (FCT, Portugal). CJM is also supported by an FCT Research Professorship, contract reference IF/00064/2012, funded by FCT/MCTES (Portugal) and POPH/FSE (EC).
\end{acknowledgments}

\end{document}